 \documentclass[apj,numberedappendix]{emulateapj}

\usepackage{color}

\def\simlt{\lower.5ex\hbox{$\; \buildrel < \over \sim \;$}}
\def\simgt{\lower.5ex\hbox{$\; \buildrel > \over \sim \;$}}

\def\beq{\begin{equation}}
\def\eeq{\end{equation}}
\def\ba{\begin{eqnarray}}
\def\ea{\end{eqnarray}}

\def\Sect{{\rm Section}}

\def\Eq{Equation}
\def\Eqs{Equations}

\def\M{{\cal M}}
\def\Lsd{L_{\rm sd}}
\def\sT{\sigma_{\rm T}}
\def\LHE{L_{\rm IC}}
\def\IGJ{I}
\def\NGJ{N_0}

\def\fej{\xi}
\def\labs{l_{\rm abs}}
\def\gbr{\gamma_{\rm br}}

\def\tamb{t_{\rm amb}}

\def\Nej{N_{\rm ej}}
\def\Rs{R_{\rm TS}}
\def\n{\tilde{n}}
\def\B{\tilde{B}}
\def\D{\tilde{\Delta}}
\def\Gcr{\Gamma_{\rm cr}}

\def\Gf{\Gamma_{\Delta}}
\def\Lf{L_{\rm f}}

\def\Gw{\Gamma_{\rm w}}
\def\Lw{L_{\rm w}}
\def\Bw{B_{\rm w}}
\def\Ef{E_{\rm f}}
\def\sigw{\sigma_{\rm w}}
\def\dNw{\dot{N}_{\rm w}}
\def\rL{r_{\rm L}}
\def\nuobs{\nu_{\rm obs}}

\def\eff{\varepsilon}
\def\tauf{\tau_{\rm f}}
\def\E{{\cal E}}
\def\N{{\cal N}}
\def\I{{\cal I}}
\def\Ebw{{\cal E}_{\rm bw}}
\def\EFRB{{\cal E}_{\rm FRB}}

\newbox\grsign \setbox\grsign=\hbox{$>$} \newdimen\grdimen \grdimen=\ht\grsign
\newbox\simlessbox \newbox\simgreatbox \newbox\simpropbox
\setbox\simgreatbox=\hbox{\raise.5ex\hbox{$>$}\llap
     {\lower.5ex\hbox{$\sim$}}}\ht1=\grdimen\dp1=0pt
\setbox\simlessbox=\hbox{\raise.5ex\hbox{$<$}\llap
     {\lower.5ex\hbox{$\sim$}}}\ht2=\grdimen\dp2=0pt
\setbox\simpropbox=\hbox{\raise.5ex\hbox{$\propto$}\llap
     {\lower.5ex\hbox{$\sim$}}}\ht2=\grdimen\dp2=0pt
\def\simgt{\mathrel{\copy\simgreatbox}}
\def\simlt{\mathrel{\copy\simlessbox}}

\begin{document}

\title{
A flaring magnetar in FRB 121102?
}

\author{
Andrei M. Beloborodov
}
\affil{Physics Department and Columbia Astrophysics Laboratory,
Columbia University, 538  West 120th Street New York, NY 10027
}

\begin{abstract}
The persistent radio counterpart of FRB~121102 is estimated to have 
$N\sim 10^{52}$ particles, energy $E_N\sim 10^{48}$~erg, and size
$R\sim 10^{17}$~cm. The source can be
a nebula inflated and heated by an intermittent 
outflow from a magnetar --- a neutron star powered by its magnetic 
(rather than rotational) energy. The object is young and frequently 
liberating energy in magnetic flares driven by accelerated ambipolar diffusion 
in the neutron star core, feeding the nebula and producing bright millisecond bursts.
The particle number in the nebula is consistent with ion ejecta from giant flares.
The nebula may also contain the freeze-out of electron-positron pairs $N_\pm\sim 10^{51}$
created months after the neutron star birth; the same mechanism offers an explanation 
for $N_\pm$ in the Crab nebula.
The persistent source around FRB~121102 is likely heated by magnetic dissipation 
and internal waves excited by the magnetar ejecta. 
The volumetric heating by waves explains the nebula's enormous efficiency in producing 
radio emission. The repeating radio bursts are suggested to occur much closer to 
the magnetar,
whose flaring magnetosphere drives ultrarelativistic internal shocks into the magnetar wind. 
The shocks are mediated by Larmor rotation that forms a GHz maser 
with the observed ms duration. Furthermore, the
flare ejecta can become charge-starved and then convert to electromagnetic waves.
\end{abstract}



\section{Introduction}

The repeating source of fast radio bursts FRB 121102 has been active since its discovery 
4 years ago (Spitler et al. 2016).
The detection of its persistent counterpart (Chatterjee et al. 2017) led to accurate 
localization and the discovery of its host dwarf galaxy at redshift $z=0.193$ 
(Tendulkar et al. 2017). This establishes the distance to the source, its 
persistent radio luminosity $L\approx 10^{39}$~erg/s, and the ms burst energies 
$\sim 10^{38}$~erg. The bursts and persistent emission are co-located -- the upper 
limit on their projected separation is 40~pc (Marcote et al. 2017), 
and the upper limit on the persistent source size is 0.7~pc.
The persistent spectrum has a sharp break at frequency $\nu\approx 10$~GHz.
A plausible age of the source is between 10 and 100~yr (Metzger et al. 2017).


\section{The persistent synchrotron source}

Synchrotron emission from particles with Lorentz factor $\gamma_e$ in a magnetic field 
$B$ peaks at frequency $\nu\approx 0.2\gamma_e^2eB/2\pi m_ec$, and the radiated 
spectral luminosity $L_\nu\propto \nu^{-\alpha}$ is related to the particle distribution over 
$\gamma_e$,
\beq
  L_\nu\approx 3\,\frac{e^3B}{m_ec^2}\,\frac{dN}{d\ln\gamma_e}.
\eeq
The observed $L_\nu\approx 10^{29}\nu_{10}^{-0.2}$~erg at $\nu<10$~GHz gives
\beq
\label{eq:NB}
  NB\sim 2\times 10^{50}{\rm ~G}.
\eeq
The 10~GHz spectral break  
(from $\alpha\approx -0.2$ to $\alpha\approx -1$)
 implies that the energy of the emitting plasma $E_N$ peaks at 
$\gamma_e\sim\gbr\approx 10^2 B^{-1/2}$ (where $B$ is in Gauss).  
Then the characteristic magnetization parameter of the source of size $R$, 
$\sigma=R^3 B^2/3E_N$, is
\beq
\label{eq:sigma}
  \sigma\approx \frac{R^3 B^2}{3N\gbr m_ec^2}
  \approx \frac{3\times 10^{54}\,B^{5/2} }{N}\,R_{17}^3. 
\eeq
\Eqs~(\ref{eq:NB}) and (\ref{eq:sigma}) now yield,
\begin{eqnarray}
   B &\sim& 0.06\,\sigma^{2/7} R_{17}^{-6/7} {\rm~G} \\
   N &\sim& 3\times 10^{51}\,\sigma^{-2/7} R_{17}^{6/7} \\
   E_N &\sim& 10^{48}\,\sigma^{-3/7} R_{17}^{9/7} {\rm ~erg}.
\end{eqnarray}

One can test the value of $R$ by looking at its implications for self-absorption and 
cooling breaks. 
The transition to efficient cooling occurs at the Lorentz factor,
\beq
  \gamma_c=\frac{6\pi m_ec}{\sT B^2 t}
  \approx 200\,t_9^{-1} \sigma^{-4/7} R_{17}^{12/7},
\eeq
where $t$ is the source age. The corresponding frequency is 
\beq
  \nu_c\approx \frac{10em_ec}{\sT^2B^3t^2}
  \approx 2\,t_9^{-2} \sigma^{-6/7} R_{17}^{18/7} {\rm GHz}.
\eeq
$R\approx 10^{17}$cm and age $t\sim 10^9$~s\,=\,30~yr 
are consistent with the cooling break being near 10~GHz.

Emission at frequency $\nu$ is dominated by electrons with energy 
$E_e\approx (\nu/10^6B)^{1/2}m_ec^2$. Equating this energy to a few times the 
brightness temperature of the source $kT_b=c^2L_\nu/8\pi^2R^2\nu^2$, one finds the 
frequency at which the observed $L_\nu$ must become self-absorbed,
\beq
   \nu_{\rm abs}\approx 2\,B^{1/5} R_{17}^{-4/5}{\rm ~GHz}
   \approx 1 \,\sigma^{2/35} R_{17}^{-34/35} {\rm ~GHz}.
\eeq
$R\approx 10^{17}$~cm is consistent with self-absorption being marginally important 
at  $\nu \sim 1$~GHz.


\section{Origin of the nebula}

The possibility of FRB association with neutron star activity was discussed in previous works
(e.g. Lyubarsky 2014; Katz 2016; Kashiyama \& Murase 2017; Dai et al. 2017).
Our estimate for the nebula energy $E_N\sim 10^{48}$~erg, the sporadic ms activity at 
its center, and the probable age $t\sim 10^9$~s, point to a young bursting magnetar,
whose older counterparts (ages $\sim 10^{11}$~s) are observed in our galaxy 
(Kaspi \& Beloborodov 2017). They have magnetic dipole moments 
$\mu\sim 10^{33}$~G~cm$^3$, hidden internal fields
$B_\star\sim 10^{16}$~G, and energies 
$E_\star\sim (R_\star^3B_\star^2/6)\sim 2\times 10^{49}B_{\star,16}^2$~erg. 
Magnetars generate multiple bursts of different energies, which cluster in time. 
Such intermittent activity is typical for evolving magnetic fields (cf. solar flares).

The magnetar spindown power decreases with time,\footnote{This standard estimate for 
   $L_{\rm sd}$ (with the braking index of 3) neglects 
   variations of $\mu$ in active magnetars and may be missing a numerical factor of a few.}
\beq
\label{eq:Lsd}
   L_{\rm sd}\approx \frac{c^3\I^2}{4\mu^2t^2}\approx 7\times 10^{36}\,\mu_{33}^{-2}t_9^{-2}
   {\rm ~erg/s},
\eeq
where $\I\approx 10^{45}$~g~cm$^2$ is the stellar moment of inertia.
At $t\sim 10^9$~s the expected  $L_{\rm sd}$ is two orders of magnitude below the 
observed radio luminosity of the nebula. 

However, the magnetar is capable of releasing its magnetic energy with a much higher rate,
\beq
   L\sim \frac{E_\star}{\tamb} \sim 10^{40}\, E_{\star,49}\,t_{\rm amb,9}^{-1} {\rm~erg/s}.
\eeq
The magnetic energy is expelled from the magnetar core due to ambipolar diffusion 
on a timescale $\tamb$.
It was previously estimated as $\tamb\sim 10^{11}$~s (Thomson \& Duncan 1996),
but recent work suggests a much shorter timescale (Beloborodov \& Li 2016).
It is controlled by proton friction against neutron liquid and sensitive to the core
temperature, which can be calculated self-consistently. 
For minimum (modified URCA) cooling, 
$\tamb\sim 10^2\,(B_\star/3\times 10^{16}{\rm~G})^{-1.2}k_{-5}^{-1.6}$~yr, where 
$k\sim 2\pi/R_\star$ describes the field gradient in the core.
Besides an ultrastrong $B_\star>10^{16}$~G, $\tamb$ may be reduced by enhanced 
neutrino cooling due to Cooper pairing of neutrons or direct URCA reactions;
the latter are activated in sufficiently massive neutron stars.
 
The magnetar was born in a supernova explosion ejecting mass $M\sim 10^{34}$~g.
The ejecta with current ballistic expansion speed $V$ 
has mass density $\rho\sim(3M/4\pi V^3t^3)$. The magnetar activity
with average power $L$ inflates a pressure bubble of radius $R$ inside the ejecta,
\beq
\label{eq:R}
   R\sim (3\epsilon LV^3t^6/M)^{1/5} 
   \approx 10^{17}\, \epsilon^{1/5} L_{40}^{1/5}V_9^{3/5}t_9^{6/5} {\rm cm},
\eeq
where $\epsilon<1$ is the power reduction factor due to radiative losses in the nebula.
This rough estimate suggests that the radio source with the estimated energy density 
$U\sim 3E_N/4\pi R^3\sim 10^{-4}$~erg/cm$^3$ is consistent with
$\epsilon L\sim 10^{39}$~erg/s and $R\sim 10^{17}$~cm.
The bubble inside ejecta with realistic $V(r)$ and $\rho(r)$ will require 
detailed calculations. The inner edge of the nebula (the wind termination shock) is at
\beq
  \Rs\approx (3L/4\pi cU)^{1/2}\approx 3\times 10^{16}\,L_{40}^{1/2}\,U_{-4}^{-1/2} {\rm cm}.
\eeq

The obtained number of particles $N\sim 10^{52}$ cannot
be supplied by the usual mechanism invoked for pulsar wind nebulae (PWN).
Pulsars create $e^\pm$ pairs with rate $\dot{N}_\pm=2\M\IGJ/e$
where $\M$ is $e^\pm$ multiplicity, and $\IGJ=\mu\Omega^2/c$ is the electric current 
circulating through the magnetosphere rotating with rate $\Omega$.
Then the $e^\pm$ number ejected over the spindown time $t$ is
\beq
\label{eq:NGJ}
   N_\pm\sim \dot{N}_\pm t\sim \M \NGJ, \;\;\; \NGJ\sim \frac{c^2\I}{e\mu}
   \sim 2\times10^{42}\,\mu_{33}^{-1}.
\eeq
insufficient for the nebula of FRB~121102.

\subsection{Pair freeze-out}

The nebula can be loaded with $e^\pm$ pairs at an early expansion stage, when it 
was mainly powered by $L_{\rm sd}\propto t^{-2}$ (\Eq~\ref{eq:Lsd}).
The early $e^\pm$ creation was noted in the context of superluminous 
supernovae (Metzger et al. 2014); below we estimate the number of pairs that 
survive annihilation and remain in the nebula. 

The power $\Lsd$ is delivered to the termination shock 
and converts to high-energy particles, which are quickly cooled through 
inverse Compton (IC) and synchrotron emission. The IC cooling is accompanied by 
$e^\pm$ cascade when the nebula has a large ``compactness parameter''
$\ell=\sT \LHE /R m_ec^3>1$, where $\LHE=f\Lsd$
is the power fraction deposited into the IC cascade.
At the early times, the nebula radius $R(t)$ (\Eq~\ref{eq:R}) should be estimated
with $L=\Lsd$ and $\epsilon\sim 1$, which yields
\beq
\label{eq:l}
  \ell(t)\approx 2f\mu_{33}^{-8/5}V_9^{-3/5}\, t_6^{-14/5}.
\eeq
When $\ell>10$, a ``saturated'' pair cascade occurs, converting $Y\sim 10$\% of 
energy into $e^\pm$ rest mass (Svensson 1987). The pair yield $Y$ decreases
at $\ell<10$; it depends on the self-consistent spectrum of soft radiation that absorbs 
the IC photons, converting them to $e^\pm$ pairs.
Below we use a rough estimate $Y\sim 0.1(\ell/10)$.

The annihilation balance in the nebula gives
\beq
\label{eq:ab}
  \frac{3Y\LHE}{4\pi R^3 m_ec^2}=\frac{3}{4}n_+n_-\sT c
  \quad\Rightarrow\quad Y\ell=\frac{\pi}{4}\tau_\pm^2,
\eeq
where $\tau_\pm\equiv \sT R n_\pm$. This balance is still (marginally) satisfied at the
freeze-out transition,
when the expansion time $R(dR/dt)^{-1}\approx t$ becomes equal to the annihilation time
$t_{\rm ann}=8/3n_+\sT c$. Thus, at freeze-out time $t_\pm$ we have both \Eq~(\ref{eq:ab}) 
and
\beq
 \label{eq:fo}
    \tau_\pm=\frac{16}{3c}\,\frac{dR}{dt}\approx 0.014 \,\mu_{33}^{-2/5} V_9^{3/5} t_6^{-1/5}.
\eeq
Equations~(\ref{eq:l})-(\ref{eq:fo}) can be solved for $\tau_\pm$, $t_\pm$, and $\ell(t_\pm)$ 
\begin{eqnarray}
   \ell_\pm &\approx& 0.1\, f^{-1/13}\mu_{33}^{-4/13} V_9^{9/13} \\   
    t_\pm &\approx& 3\times 10^6\, f^{5/13} \mu_{33}^{-6/13}V_9^{-6/13} {\rm ~s}
\end{eqnarray}
The pair yield at freeze-out $Y_\pm\sim 10^{-2}\ell_\pm$ then determines the number 
of $e^\pm$ that remain in the nebula,
\beq
\label{eq:Nf}
   N_\pm\approx Y_\pm\,\frac{f\Lsd t_\pm}{m_ec^2}
   \approx  2\times 10^{51}\,f^{7/13}\mu_{33}^{-24/13}V_9^{15/13}.
\eeq

\subsection{Matter ejection in magnetar flares}

Magnetars eject plasma during their giant flares.
The most powerful flare observed to date occured in SGR~1806-20
in 2004. It radiated $E_\gamma\sim 2\times 10^{46}$~erg in gamma-rays 
(Palmer et al. 2005) and was followed by radio afterglow emitted by mildly relativistic 
ejecta (Gaensler et al. 2005). Granot et al. (2006) estimated a lower limit on the ejecta 
mass $M_{\rm ej}>3\times 10^{24}$~g, which corresponds to a minimum number of
ejected ions $N_{\rm ej}\sim 2\times 10^{48}$.
The ejecta are dominated by electron-ion plasma
(annihilation limits $e^\pm$ ejection, see \Sect~5).

The lower limit on $\Nej$ in SGR~1806-20 corresponds to 
$N_{\rm ej}/E_\gamma>10^2$~particles per erg emitted in gamma-rays. An upper limit is 
set by the condition that the flare energy per proton exceeds the gravitational 
binding energy $GM_\star m_p/R_\star\approx 10^{-4}$~erg. This gives
\beq
  10^2< \fej\equiv\frac{N}{E}<10^4 {\rm ~erg}^{-1}.
\eeq
An active magnetar, releasing a total of $E\sim 10^{49}$~erg in flares, supplies a large 
number of particles to the nebula, 
\beq
\label{eq:lift}
   N \sim 10^{52}\,\left(\frac{\fej}{10^3/\rm erg}\right)\, E_{49}.
\eeq


\section{Heating of the nebula}

Observations of SGR~1806-20 show that giant flares eject chunks of matter with speeds 
$v\sim 0.3-0.7c$. Velocity dispersion within the ejecta $\delta v\sim v$ implies a spread in 
its time of arrival to a radius $r$: $\delta t(r) \sim r/\delta v\sim 10^6\,r_{16}$~s.
Because of this spreading, frequent impulsive flares 
create a quasi-continual wind before it reaches the termination shock
(if the flare rate exceeds $\sim 10^{-6}\,\Omega_b^{-1}$~s$^{-1}$, where $\Omega_b$ 
is the ejecta solid angle).
The baryon-rich ejecta, which initially carry a fraction of the magnetar power, 
interact and mix with the more energetic high-$\sigma$ flow from subsequent flares
before they reach the nebula. 
This creates a wind with average energy per particle $\sim E/N\sim 1$~GeV 
(corresponding to $\xi\sim 10^3$~erg$^{-1}$ in \Eq~\ref{eq:lift}).
The variable wind is reheated by internal shocks before the termination shock.

The hot, marginally supersonic, wind is decelerated by a weak termination shock, or 
by a smooth pressure gradient, and joins the nebula.
The deceleration of the variable wind occurs with variable compression, which implies
a persistent source of sound waves with frequencies $\nu\simgt c_s/\Rs$ 
and wavelengths $\lambda=c_s/\nu$.

These waves will propagate through the nebula and dissipate over time.  
They dissipate due to heat conduction and viscosity, controlled by the 
particle diffusion coefficient $D\sim l c$, where $l$ is the particle mean free path. 
The waves dissipate on a scale $\labs>\lambda$ if $D<c\lambda$, 
\beq
   \labs\sim\frac{\lambda^2}{l}, \qquad l<\lambda.
\eeq
Particles  follow the magnetic field lines in the nebula, as their Larmor radii are small, 
$r_{\rm L}=\gamma_e m_ec^2/eB\sim 10^8$~cm.
Their effective $l$ does not exceed the correlation length of the magnetic field, 
a fraction of the nebula radius. The magnetic field structure may be complicated by 
surviving high-$\sigma$ stripes from the flare ejecta.

A plausible $l<\Rs$ allows 
sound waves with $\lambda>(l\Rs)^{1/2}$ to propagate and transport energy through
the nebula. In the opposite case, $l\simgt \Rs$, 
effective energy transport occurs as well, now due to heat diffusion. 
We conclude that wave excitation by the variable 
magnetar wind provides efficient volumetric heating of the nebula.
It was not studied before in normal PWN (Gaensler \& Slane 2006), where
most of the dissipated energy is deposited near the steady termination shock into 
a small number of high-energy particles radiating at high frequencies.

The released heat is partitioned between ions and electrons, with an energy distribution 
which may be broadened by magnetic reconnection events and stochastic 
particle acceleration by the turbulence. At low energies, the electron distribution is
affected by synchrotron self-absorption, which must create a low-energy break.
Electrons with energies exceeding $\gamma_c m_ec^2$
(which happens to be comparable with the mean particle energy $\sim 0.1$~GeV)
will cool faster than the age of the nebula, increasing its radiative losses.
The characteristic $B\sim 0.03$~G, the volumetric heating sustaining 
$\gamma_e\sim 300$, and the proximity of $\gamma_e$ to $\gamma_c$ are what 
makes the nebula so efficient in radiating its energy in synchrotron radio waves.

If the magnetar releases energy in rare energetic flares, the ejecta  do not mix 
into a continual wind at $r<\Rs$. Then the leading high-$\sigma$ part of the flare ejecta 
preserves its high power and suddenly applies a 
huge pressure to the nebula, pushing out the termination shock. The 
impact power is $\Lf=\Ef/\tauf$, where $\Ef$ is the flare energy and $\tauf$ is its duration.
The dynamics of such an impact and accompanying radiation was discussed
by Lyubarsky (2014) and Murase et al. (2016).
A sufficiently high $\Lf$ launches a forward shock into the nebula
with Lorentz factor $\Gamma_0\approx (L/4\pi r^2 cU)^{1/4}$ 
(where $U\sim 10^{-4}$~erg/cm$^3$ is the nebula energy density)
while the reverse shock continues to propagate through the ejecta. 
The reverse shock crosses the ejecta after time $\tau_0=\Gamma_0^2\tauf$, 
and the forward shock begins to decelerate.
Its Lorentz factor $\Gamma$ decreases with approximate energy conservation, 
\beq
  4\pi r^2 c\tau U\Gamma^2\approx \Ef, \qquad 
  \Gamma(\tau)\approx \Gamma_0\left(\frac{\tau}{\tau_0}\right)^{-1/2}.
\eeq
The shock becomes a sound wave when $\Gamma$ decreases to $\Gamma_s\sim 1$
(unless the upstream $\sigma\gg 1$; then $\Gamma_s\approx \sigma^{1/2}$). 
This occurs after time $\tau_s=\tau_0(\Gamma_0/\Gamma_s)^2$, which determines 
the length traveled by the shock,
\beq
  l_s=c\tau_s\sim 10^{16}\,E_{\rm f,45}\Gamma_s^{-2}U_{-4}^{-1}  
  R_{\rm TS,16}^{-2} {\rm ~cm}.
\eeq
The shocked plasma cools on timescale $t_{\rm syn}\propto\Gamma^{-2}\propto \tau$
and loses energy fraction $\tau/t_{\rm syn}=\tau_s/t_{\rm syn}^{\rm neb}\ll 1$
where $t_{\rm syn}^{\rm neb}\sim 10^9$~s is the cooling time in the nebula
ahead of the shock.
At times $\tau>\tau_s$, most of the impact energy $\Ef$ is carried by a strong sound 
wave. Its wavelength is $\lambda\sim l_s$ and its initial amplitude is $\sim 1$.
Its heating effect is similar to that of sound waves 
generated by the mixed mildly relativistic wind described above.


\section{Fast radio bursts}

A synchrotron maser must form at the termination shock of a pulsar wind 
(Gallant et al. 1992). Lyubarsky (2014) proposed that FRBs could be emitted 
by a similar maser in a magnetar wind nebula, when its power is suddenly boosted by a 
giant flare. A shortcoming of the model is the assumed flare energy $\Ef\sim 10^{48}$~erg 
and the huge power $\Lf\sim 10^{52}$~erg/s required to push the nebula shock to 
$\Gamma_0\sim 10^4$ --- otherwise the Doppler-compressed FRB duration 
exceeded 1~ms (the minimum duration of emission arriving from a flashing spherical shock
is $r/\Gamma_0^2 c$). In addition, following a flare the termination shock cannot recover 
quicker than $\Rs/c$, contradicting the short times between bursts in FRB~121102.
Below we suggest an alternative scenario, with moderate $\Ef\sim 10^{44}$~erg and 
$\Lf\sim 10^{47}$~erg/s, which is consistent with the large number of bursts from 
FRB~121102.

Magnetars produce flares when their magnetospheres are over-twisted (Parfrey et al. 2013). 
Flares on magnetic field lines extending to $r_0\sim 10^7$~cm release 
$\Ef\sim \mu^2/r_0^3\sim 10^{45}\mu_{33}^2r_{0,7}^{-3}$~erg. At the flare onset 
a magnetic island of size $\sim r_0$ disconnects 
and accelerates away from the star on timescale $r_0/c$,
creating a shell of thickness $\Delta\sim r_0$ and carrying energy 
$\E$ comparable to $\Ef$. 

This first ejected shell is unlikely to be significantly polluted 
by baryons from the magnetar surface. 
The magnetospheric twist that triggered the flare was supported by electric current 
$I_{\rm tw}\sim c\mu/r_0^2$, which gives a minimum number of $e^\pm$ in the flare 
region $\N\sim \M (I_{\rm tw}/e)(r_0/c)$, with an expected $e^\pm$ multiplicity
$\M\sim 10^2$ (Beloborodov 2013). 
Thus, $\N_{\min}\sim \M \mu/er_0\sim  2\times 10^{37}\,\M_2\,\mu_{33}\,r_{0,7}^{-1}$.

More pairs are loaded if the shell ejection went with partial dissipation of $\E$, 
creating a thermal fireball with equipartition temperature $T_0$ 
($aT_0^4\sim \E/r_0^3$). The  expanding fireball cools from $kT_0\sim 200$~keV to 
$e^\pm$ freeze-out $kT_\pm\sim 20$~keV (Paczynski 1986)  
after expansion to $R_\pm \sim 10^8$~cm and acceleration to Lorentz factor
$\Gamma_\pm\sim 10$.
The freeze-out occurs when $\tau_\pm\sim n_\pm\sT R_\pm/\Gamma_\pm^2\sim 1$ 
which gives $\N_{\max}\sim \Gamma_\pm^2R_\pm r_0/\sT\sim 10^{41}$.

\subsection{Internal shocks}

Energy per particle in the leading $\Delta$-shell ejected by the flare is high,
$\eta_\Delta=\E/\N m_ec^2\sim 10^{11}\,\E_{44}\N_{39}^{-1}$.
As the shell expands, its energy remains concentrated within 
radial thickness $\Delta\sim r_0$ while its Lorentz factor grows as 
$\Gf\sim (\eta_\Delta r/r_0)^{1/3}$ (Lyutikov 2010; Granot et al. 2011).
The fast $\Delta$-shell drives a blast wave into the pre-flare magnetar wind. 

The blast Lorentz factor $\Gamma$ is given by pressure balance,
\beq
\label{eq:G}
   \Gamma\approx \Gw \left(\frac{\Lf}{\Lw}\right)^{1/4}
     =10^4\,\Gamma_{\rm w,2} \left(\frac{L_{\rm f,47}}{L_{\rm w,39}}\right)^{1/4},
\eeq
where $\Lf\sim \E c/\Delta= 3\times 10^{47}\E_{44}\Delta_7^{-1}$~erg/s,
$\Lw$ is the wind power, and $\Gw$ is the wind Lorentz factor. 
The pre-flare $\Lw$ is likely far above the nominal $\Lsd\sim 10^{37}$~erg/s, 
as the twisted magnetosphere was 
inflated and its dipole moment $\mu$ was increased (Parfrey et al. 2013). 
 An unknown parameter of the pre-flare wind is 
\beq
   \sigw=\frac{\Lw}{\Gw m_ec^2 \dNw}> 1, 
\eeq
where $\dNw$ is the particle outflow rate. The enhanced $\Lw$ may
involve enhanced $e^\pm$ loading;
therefore $\sigw$ may be much lower than in ordinary pulsars.

Energy transferred from the $\Delta$-shell to the blast wave grows with radius, 
\beq
   \Ebw(r)\approx \frac{r}{2\Gamma^2\Delta}\,\E
       \approx \frac{r}{2c}\left(\Lf\Lw\right)^{1/2}  \qquad (r<2\Gamma^2\Delta).
\eeq
Most of the transferred energy is stored in the swept-up wind magnetic field, and
a fraction $\sigw^{-1}$ is deposited into the shocked wind plasma. 
The wind magnetic field is transverse to the radial direction and the shock  
is mediated by Larmor rotation. The shocked particles gyrate in the fluid frame
with Lorentz factor $\gamma_e\sim \Gamma/\Gw$ and Larmor radius
$\rL\sim \Gw m_ec^2/e\Bw$, where $\Bw=(\Lw/cr^2)^{1/2}$ is the pre-shock
magnetic field measured in the lab frame.
The gyrating particles form the synchrotron maser, and a fraction of their 
energy $\eff\sim 10^{-2}$ converts to semi-coherent electromagnetic waves 
(Gallant et al. 1992) with a characteristic observed frequency
\beq
\label{eq:nuobs}
   \nuobs\sim \frac{\Gamma\,c}{2\pi \rL}
     =\frac{e\,(\Lf\Lw)^{1/4}}{2\pi m_ec^{3/2}\,r}
      \approx \frac{3{\rm ~GHz}}{r_{13}}(L_{\rm f,47}L_{\rm w,39})^{1/4}.
\eeq
The coherent radiation has energy $\EFRB\sim \eff\,\sigw^{-1}\Ebw(r)$,
\beq
\label{eq:EFRB}
  \EFRB\sim 10^{39}\, r_{13}\,\varepsilon_{-2}\,\sigw^{-1}\,\Gamma_{\rm w,2}^{-2}
    (L_{\rm f,47}L_{\rm w,39})^{1/2}{\rm ~erg},
\eeq
and observed duration
\beq
   \tau_{\rm obs}\sim \frac{r}{\Gamma^2c}
     \approx 3\times 10^{-6}\,r_{13}\,\Gamma_4^{-2} {\rm ~s}.
\eeq
A lower $\Gamma$ gives a longer duration; then $\tau_{\rm obs}$
can become related to the $\Delta$-shell thickness and the observed time of its energy 
transfer to the blast wave, $\Delta/c\sim 1$~ms.

As the shock expands to radius $r$ it sweeps up wind material that was emitted
by the magnetar during a small time $\delta t$ just before the flare:
$\delta t \sim r/c\Gw^2\sim 3\,r_{15}\Gamma_{\rm w,2}^{-2}$~s.
This time could exceed the magnetar spin period $P$, which imprints periodicity 
on the swept-up wind; then the FRB emission is modulated with period
$(\Gw/\Gamma)^2P$.


\subsection{Charge starvation}

The ideal (force-free) picture of the expanding $\Delta$-shell is valid as long as it 
sustains current density $\tilde{j}$ demanded by the gradient of its magnetic field 
$\B/\D=\Gf^{-2}(B/\Delta)$ (tilde indicates the shell rest frame). 
The maximum possible $\tilde{j}$ is $ec\n_\pm$ where $\n_\pm=n_\pm/\Gf$ is
the proper density and $n_\pm=\N/4\pi r^2\Delta$. It falls short of the required 
current $\tilde{j}\sim (c/4\pi)\B/\D$ if
\beq
\label{eq:Gcr}
  \Gf<\Gcr(r)=\frac{r}{e\N}\left(\frac{\E}{\Delta}\right)^{1/2} 
                   \approx 400\, r_{14}\,  \N_{39}^{-1}\E_{44}^{1/2}\Delta_7^{-1/2},
\eeq
where $B^2r^2\approx \E/\Delta$ was used. A freely expanding $\Delta$-shell with
$\Gf\sim (\eta_\Delta r/\Delta)^{1/3}= 10^6 r_{14}^{1/3}\E_{44}^{1/3}\N_{39}^{-1/3}\Delta_7^{-1/3}$ 
will not experience charge starvation.

However, the initial free expansion is inevitably followed by
deceleration. At the latest, this occurs at the termination shock $\Rs$.
Deceleration at smaller $r$ occurs when the
$\Delta$-shell runs into the slow baryonic ejecta tail from a previous magnetar flare. 

The deceleration will trigger charge starvation, and
the $\Delta$-shell energy will partially convert to vacuum electromagnetic waves.
The observed duration of this event is
\beq
   \tau_{\rm obs}\sim \frac{r}{\Gcr^2 c}
   \sim 10^{-3}~r_{14}^{-2}\N_{39}^2\,\E_{44}^{-1} {\rm~s}.
\eeq
The main wave frequency $c/\Delta\sim 1$~kHz is too low to be interesting,
however a fraction $f$ of wave power might emerge at GHz frequencies. 
A GHz burst with energy $\sim 10^{38}\E_{44}$~erg would require $f\sim 10^{-6}$.


\section{Discussion}

The energy and number of particles $N \sim 10^{52}$ in the nebula of FRB~121102 are
consistent with magnetar ejecta, based on the observations of SGR~1806-20. 
This explanation of $N$ implies that the 
nebula is made of $\sim 3\times 10^{-6}M_\sun$ of the magnetar crustal material.
The electron-ion nebula could create Faraday rotation for linearly polarized FRBs. 
A correlation length of the magnetic field $l\simlt R$ gives,
with the nebula parameters estimated in \Sect~2, a modest rotation measure 
$\sim 10(l/R)$~rad/m$^{2}$, depending on the electron 
distribution at low $\gamma_e<100$.

The frequent bursts of FRB~121102 imply many 
flares, many more than observed from the local magnetar population.
This may not be surprising, as the local magnetars are about hundred times older and 
mostly dormant (Kaspi \& Beloborodov 2017). 
Local magnetar wind nebulae (MWN) are hardly detectable because
of their age and weaker activity.
The only observed MWN, in Swift~J1834.9-0846 (Yunes et al. 2016),
is consistent with a magnetar flare origin (Granot et al. 2017). 
The evolution of old magnetars is likely driven by the (relatively slow)
Hall drift of their crustal magnetic fields rather than ambipolar diffusion in the core. 

Besides the young age, the hyper-active FRB~121102 probably has
an unusual progenitor. The nebula size $R\simgt 10^{17}$~cm is 
consistent with a hyper-energetic supernova shell accelerated to 
$V\simgt 10^9$~cm/s. The shell energy
$MV^2/2\sim 10^{52}$~erg may come from the magnetar birth 
with rotational energy $\I\Omega^2/2\sim 2\times 10^{52}P_{\rm ms}^{-2}$~erg, 
where $P\sim 1$~ms is the spin period. 
By contrast, local magnetars have supernova shells with energies 
$\sim 10^{51}$~erg (Vink \& Kuiper 2006).

Ultra-fast initial rotation likely generates exceptionally strong magnetic fields
(Duncan \& Thompson 1992).
This implies faster ambipolar diffusion in the magnetar core, resulting 
in enhanced energy release through magnetic flares.
In addition, the magnetar might have a large mass, which can enhance its 
neutrino cooling, further accelerating the ambipolar drift (Beloborodov \& Li 2016). 
The progenitor of FRB~121102 likely had a
low metallicity, consistent with the rare type of its host galaxy;
such hosts are also typical for GRBs and hydrogen-poor superluminous supernovae 
(SLSN), suggesting a connection (Metzger et al. 2017). 

The frequent bursting partially compensates for the low birth rate of objects like 
FRB~121102 and allows them to contribute to the observed FRB rate.
Magnetars produced by ordinary progenitors are less active but can emit   
FRBs by the same mechanism. The non-detection of FRB from SGR~1806-20 giant flare 
(Tendulkar et al. 2016) might be explained by a limited solid angle of the blast wave 
with sufficiently high $\Gamma$. Its pre-flare wind may be weaker 
compared with hyper-active younger magnetars,
and a low $\Lw$ makes the frequency $\nuobs$ low (\Eq~\ref{eq:nuobs}). 
Furthermore, collisions between ejecta from rare flares do not occur.

The magnetar should be spun down to $P\sim 2\,\mu_{33}\,t_9^{1/2}$s
and its current $\Lsd$ is small. However, $\Lsd$ was high in the past. As a result,
at an age $t\sim 1$~month, the energetic compact nebula must have experienced the 
freeze-out of $e^\pm$ pairs with a significant $N_\pm$ (\Eq~\ref{eq:Nf}). 
The relict pairs are likely mixed with later ejecta from flares.

The same calculation of freeze-out $N_\pm$ should apply to ordinary PWN
with $V\sim 10^{8}$~cm/s and $\mu\sim 10^{31}$~G~cm$^3$, which gives a large
$N_\pm$. This offers a solution to the old puzzle of $\simgt 10^{51}$ 
low-energy particles inferred from radio observations of the Crab nebula (Shklovskij 1968).
The relict pairs may be reheated by later magnetic dissipation.

Internal shocks described in \Sect~4 generate a train of multiple ms bursts at small radii,
well before the ejecta reach the nebula. 
The clustering of bursts in time is of particular interest; it implies more efficient collisions 
between the flare ejecta. FRB~121102 has demonstrated multiple bursts separated by 
minutes to hours, and an intermittent pattern of enhanced magnetar activity.

In the picture suggested in this Letter, two factors should control the future evolution 
of FRB~121102: the evolution of the magnetar flaring activity and the 
ballistic expansion of the supernova shell confining the nebula. Both should evolve with 
age (likely $\sim 10^9$~s timescale). 
Heating of the nebula by a single flare may not strongly boost its 
persistent radio emission; a period of enhanced magnetar flaring can create a stronger
heating impact accumulating on the sound crossing time $R/c_s\sim 0.5$~yr.

\acknowledgements
I am grateful to B. Metzger, Y. Levin, C. Lundman, L. Sironi, and J. Zrake 
for discussions and the referee for comments on the manuscript. 
This work was supported by a grant from the Simons Foundation 
(\#446228).

 \end{document}